# Level-crossing and modal structure in microdroplet resonators

Sarah T. Attar[1*], Vladimir Shuvayev[2], Lev Deych[2,3], Leopoldo L. Martin[1] and Tal Carmon[1]

[1]*Department of Mechanical Engineering, Technion-Israel Institute of Technology, 3200003 Haifa, Israel*
[2]*Physics Department, Queens College of CUNY, Flushing, NY 11367*
[3]*The Graduate Center of CUNY, 365 5th Ave, New York, NY 10016*
*\* stattar@tx.technion.ac.il*

**Abstract:** We fabricate a liquid-core liquid-clad microcavity that is coupled to a standard tapered fiber, and then experimentally map the whispering-gallery modes of this droplet resonator. The shape of our resonator is similar to a thin prolate spheroid, which makes space for many high-order transverse modes, suggesting that some of them will share the same resonance frequency. Indeed, we experimentally observe that more than half of the droplet's modes have a sibling having the same frequency (to within linewidth) and therefore exhibiting a standing interference-pattern.

## 1. Introduction

An optical whispering-gallery resonator [1] confines circulating light by total internal reflection on its spheroidal interface. This boundary must therefore border between an internal high refractive-index region and an external low refractive-index region. The spheroid interface can be a solid-gas interface [2-4], a solid-liquid [5, 6], a liquid-gas [7-13], or a liquid-liquid one [14-16].

In this paper we will focus on liquid-liquid micro-resonators. Liquid resonators are easy to fabricate. They are exceptionally smooth, and therefore minimize scattering losses. Also, liquid resonators enable phenomena that cannot occur in solid ones, including the propagation of capillary waves [11]. Recently, liquid resonators were miniaturized and fiber coupled [9, 12, 16], making them attractive for various applications and fundamental studies.

Here, we will map the optical modes of a micro-droplet resonator, and we will show crossover between many of the droplet's modes [17, 18].

In more details, the eigenfunctions of an optical whispering-gallery mode can be indexed by 3 quantum numbers: *n*, *l*, and *m*, which represent the radial, polar, and azimuthal modal-order. The circulation direction is given by the sign of *m*: positive for clockwise circulating mode and negative for counter clockwise circulating mode. As will be shown below, more than half of the modes in our droplet resonator have a "sibling" at a distance smaller than the resonance linewidth. As a result, we can pump two co-circulating modes with different *n*, *l*, *m* indices, with a single-frequency laser [17, 18].

Such pairs of modes typically exhibit various interference patterns along the circulation direction of light. The interference pattern does not move since its phase velocity, $\delta\omega/\delta k$, is zero. Therefore, the pattern can be referred to as "stopped light" [17]. In a sphere, these degenerated modes were viewed as a standing zigzag patterns along the equator [17], made by the interference between the two modes. In a toroid, such degenerated modes gave rise to a standing flower-shape interference [18]. In this flower shape, the number of petals along the azimuthal direction describes the *m* difference and the number of petal-circles describes the *n* difference.

The phenomenon is called "level-crossing" [18] since the frequencies of the modes will typically cross each other upon changing one of the parameters (e.g. refractive index). Level-crossing can be avoided (mode veering) when the modes differ only by their *n*, *l* indices and non-avoided when *m* also differs, as explained in more details in [18]. Due to the large transverse cross-section of the droplet, there are more high-transverse-order modes than, for example, in a toroid, and therefore an abundance of level crossings are expected.

Degeneracy between clockwise and counterclockwise modes is removed in the presence of nanoparticles [19-21], which may manifest experimentally either as mode splitting [22], or, if the splitting is as smaller than the width of the modes, as a frequency shift [5]. Additional degeneracy due to level-crossing is expected to enhance this effect, and results in a more sensitive detection. An additional increase in sensitivity specific for liquid resonators comes from the fact that an analyte can, in principle, be encapsulated by the resonator, and therefore, resides in the region of maximum intensity of the field.

## 2. Experimental setup

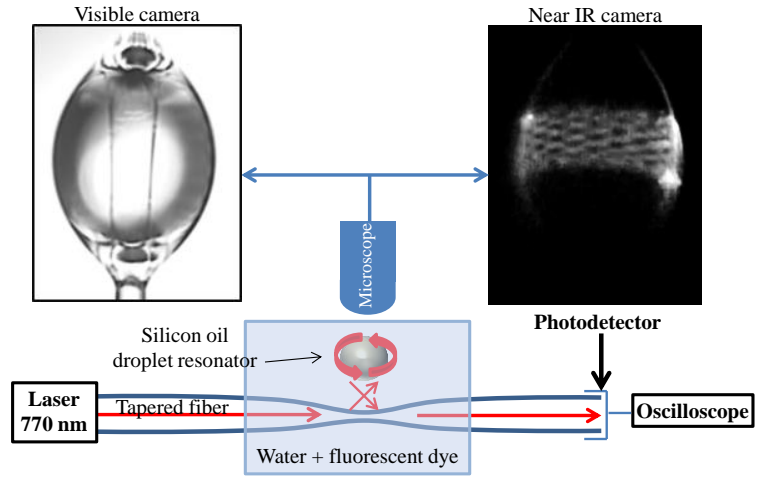

Fig. 1. Experimental setup. The light is coupled from a tapered fiber to circumferentially circulate in a microdroplet resonator. The tapered fiber and the resonator are placed in a solution of fluorescent dye in water, which allows us to see the modes in the droplet using a near IR camera.

We fabricate our resonator by dipping a silica cylinder in silicon oil. When the cylinder is taken out of the oil, a droplet is attached to its end. Fusing the extremity of the cylinder into a microsphere helps in pinning the droplet at the end of the cylinder while prohibiting its axial motion.

As one can see in Fig. 1, our experimental setup includes a micro-droplet resonator and its aquatic surrounding. The water environment contains a 3 μmol/L solution of ADS780WS fluorescent dye. This fluorescent material absorbs light at the wavelength of the optical mode (770 nm) and incoherently emits at 810 nm. The emission from the fluorescent material will be used to map the optical mode. A filter is placed between the camera and the microscope to block any scattered light from the laser, and to selectively transmit the light emitted from the excited dye's molecules. The advantage of the fluorescent mode mapping technique [17, 18] is that incoherent light can be imaged at high resolution (when compared to coherent light). In addition, the fluorescent light emerges from the thin evanescent region only, and therefore can be easily contained within the microscope depth-of-focus.

Optical coupling of the light into the cavity is performed via a tapered fiber [23-26] and the typical measured Q-factor in our experiments is $10^6$.

## 3. Experimental results

We couple light inside the resonator, and scan the laser's emission wavelength from 770 nm to 780 nm. Using the near IR camera, we record the shape of the modes while the laser is swept through them.

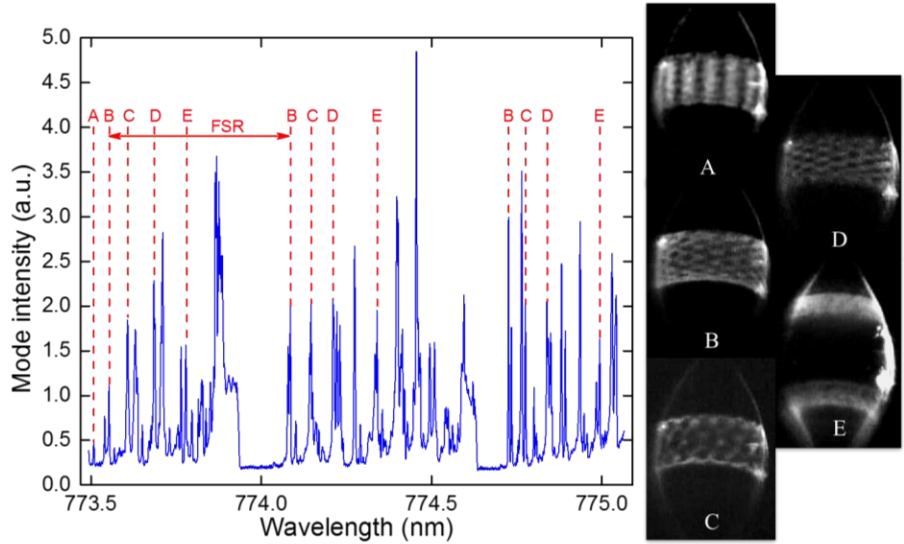

Fig. 2. Experimental results. We show the mode intensity as measured from the fluorescent emission versus laser's wavelength. Modes B-E repeats themselves every free spectral range.

As typical to such resonators [17, 18], two shapes of resonances were seen. In more details, throughout the scan we observe both single modes (Fig. 2.E), and level-crossings (Fig. 2.A-D) representing spectrally-overlapping pairs of modes. Surprisingly, we notice that about 80% of the modes have an overlapping sibling. As one can see in Fig. 2, modes shapes tend to repeat themselves every 0.6 nm. This value is in good agreement with our calculated value of the free spectral range (FSR), which is 0.59±0.01 nm.

## 4. Theory and modeling of experimental results

In order to verify that experimentally observed patterns indeed correspond to crossing modes and to assist with mode identification we carried out numerical simulations of the mode in the droplet using finite-element method FEM realized in COMSOL complimented with approximate analytical estimates. To carry out the simulations we first need to model the shape of the droplet. Using the equilibrium condition for the droplet surface [27] in the form

$$\frac{1}{r_1}+\frac{1}{r_2}=\frac{\Delta p}{\gamma}, \qquad (1)$$

where $r_1$, $r_2$ are the principal radii of curvature, $\Delta p$ is the pressure difference across the fluid interface, and $\gamma$ is the surface tension, one can find an explicit analytical shape form the droplet profile [28].

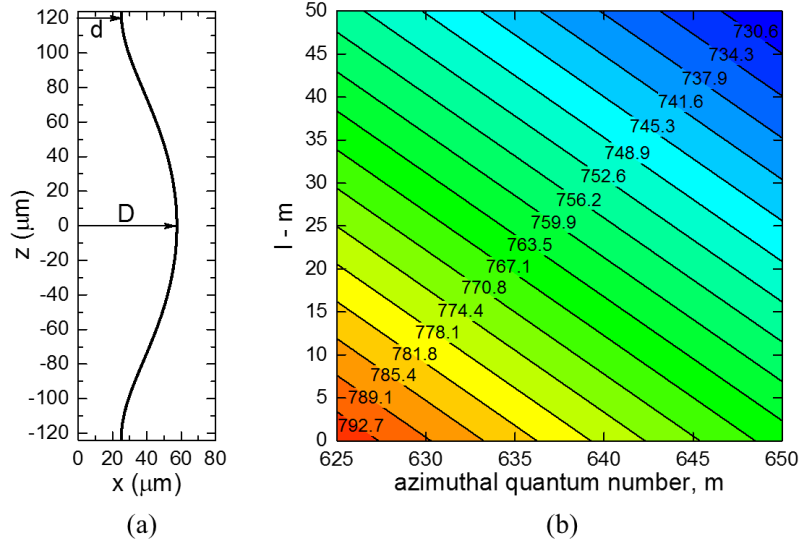

Fig. 3. (a) Profile of the droplet with equatorial radius D on the fiber with radius d. (b) A level plot of Eq. (6) corresponding to different first radial order TE resonances with wavelengths expressed in nm. One can see that multiple resonances with different values of orbital and polar numbers can appear at the chosen frequency within some linewidth.

Using a cylindrical coordinate system with its polar axis $Z$ along the axis of symmetry of the droplet (see Fig. 3(a)) the cross-section of the droplet in the $X$-$Z$ plane can be presented in the form of elliptical integrals $F(\varphi,k)$ and $E(\varphi,k)$ of the first and second order respectively:

$$z=\pm\left[ad\,F(\varphi,k)+D\,E(\varphi,k)\right]. \qquad (2)$$

Here $d$ and $D$ are radius of the fiber and the equatorial radius of the droplet, parameter $a$ depends on the contact angle $\theta$ as

$$a=\frac{D\cos\theta-d}{D-d\cos\theta}, \qquad (3)$$

while the argument $\varphi$ of the elliptic integrals is related to the coordinate $x$ of the point on the droplet surface according to

$$x^2=D^2\left(1-k^2\sin^2\varphi\right) \qquad (4)$$

with $k$ defined as $k^2=1-a^2d^2/D^2$. The equatorial plane of the droplet corresponds to $\varphi=0$ and, obviously, $x=D$. In the vicinity of the equatorial plane, the surface of the droplet can be approximated by a spheroid, in which case Eq. (2) must reduce to that of an ellipse. Indeed, considering limit $\varphi \ll 1$, we can bring this equation to the form

$$z = \pm \frac{ad + D}{D} \sqrt{D^2 - x^2}, \qquad (5)$$

which describes ellipse with large and small semiaxes $R_1 = ad + D$ and $R_2 = D$, respectively, and ellipticity parameter $e = D/(ad + D)$. Using $D = 57.5\,\mu m$, $d = 25\,\mu m$ and $a = 1$, we find that $e = 0.7$. Fitting numerically Eq. (2) by an ellipse, we find a better approximation for the parameters of the effective spheroid with $R_1 = 91.3\,\mu$m and $e = 0.63$. Assuming that the WGMs excited in our experiments do not deviate too far from the equatorial plane, we can estimate their frequencies by combining the spheroidal approximation with the asymptotic expression obtained in the geometrical optics limit. Following [29], we can present them in the form

$$n_r k_{l,m,n} R_1 = m - \alpha_n \left(\frac{m}{2}\right)^{1/3} + \frac{(2l - 2m + 1) R_1}{2 R_2} - \frac{P n_r}{\sqrt{n_r^2 - n_0^2}} + \frac{3 \alpha_n^2}{20} \left(\frac{m}{2}\right)^{-1/3} - \\ \frac{\alpha_n}{12} \left[\frac{(2l - 2m + 1) R_1^3}{R_2^3} + \frac{2 n_r^3 P (2P^2 - 3)}{\left(n_r^2 - n_0^2\right)^{3/2}}\right] \left(\frac{m}{2}\right)^{-2/3}, \qquad (6)$$

where $k_{l,m,n} = \omega_{l,m,n}/c$ is the vacuum wave number corresponding to WGM with frequency $\omega_{l,m,n}$, $n_r$ and $n_0$ are the refractive indices of the resonator and surrounding medium respectively, $\alpha_n$ is the n-th zero of the Airy function defining the radial behavior of the mode. Parameter $P$ is equal to unity for TE modes and $P = (n_0/n_r)^2$ for TM modes. In the case of spherical resonators ($R_1 = R_2$) the dependence of these frequencies on the azimuthal number $m$ vanishes and one reproduces well-known asymptotic formulas first derived in [30]. In order to illustrate the ubiquitous nature of the level crossings in non-spherical resonators we plot a level plot of Eq. (6) for TE modes of the first radial order close to the experimentally found resonances. One can see from Fig. 3(b) that multiple resonances with different values of the polar and azimuthal numbers can appear at almost the same frequency resulting in the level crossing phenomenon.

Using the analytical results as a starting point we also carried out accurate FEM simulation of the resonances using Eq. (2) to describe the shape of the drop and COMSOL Multiphysics computational platform to calculate frequencies, electromagnetic field patterns and other properties of WGMs excited in the drop. Due to the presence of the axisymmetric and reflection symmetries, simulations were limited to the 2D domain and, moreover, only half of the profile ($z \geq 0$) shown on Fig. 3(a) was used, surrounded by the correct set of the boundary conditions around the narrow region of WGM volume existing close to the surface of the drop. More details and specifics can be found, for example, in [31] and references therein, which uses "weak-form" expressions to analyze WGMs of dielectric microresonators. Numerical data from simulations with the wide range of values of $m$ and within the spectral range under interest was exported to Matlab for further analysis and visualization. This way we were able to select pairs of WGMs satisfying imposed conditions based on the experimental data. They include, but not limited to: change of *m* (which defines the number of fringes in azimuthal direction), spectral distance between modes' frequencies, vertical apparent size of the patterns, and so on. A couple of the results of the selection can be seen on Fig. 4, showing the normalized intensity of a single high-order mode (a) and the intensity of the sum of two level-crossed normalized modes (b) taken from the close vicinity to the exterior side of the drop.

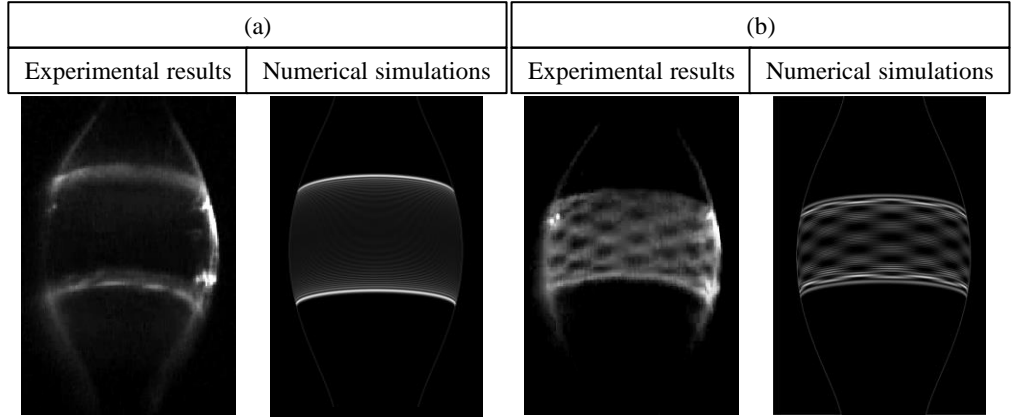

Fig 4. Experimental results and simulations for (a) a high-order mode, similar to the ($n=1$, $m=588$, $l=679$) mode, and (b) a level-crossed pair, similar to the interference between ($n=2$, $m=602$, $l=653$) and ($n=2$, $m=610$, $l=649$) modes. Numerical results were outlined to show profile of the drop.

In more details, Fig. 4(a) depicts a mode shape that is rotational symmetric along the axial direction. Our analysis shows that this mode is similar to the ($n=1$, $m=588$, $l=679$) mode. Contrary to this mode shape, the mode in Fig. 4(b) exhibits a zigzag pattern along the azimuthal direction. It is not rotational symmetrical, yet rotation by 360/8 degrees does not change the pattern. Our analysis shows that this shape is typical to the interference between ($n=2$, $m=602$, $l=653$) and ($n=2$, $m=610$, $l=649$) modes that have a sufficiently close resonance frequency yet a different *m* number. As mentioned, the 8 fringes along the azimuthal direction represent the *m* numbers' difference.

## 5. Conclusion

Experimentally mapping the resonances of a liquid micro-resonator revealed that most of the droplet's modes have a sibling at a frequency distance that is comparable to their linewidth. The interference between such mode-pairs produces various standing patterns which are called level-crossing. We also carried out numerical simulations of WGM in this system with COMSOL, using an analytical expression derived from Laplace's formula [27, 28] to describe the drop's shape. Simulations showed that the studied system possesses a large number of spectrally close modes that can produce a whole gallery of various interference patterns. On one hand, the ubiquitous nature of this effect presents a very interesting opportunity for studying various types of level-crossing phenomena, but on the other hand, makes unambiguous identification of the participating modes more difficult. The crossing modes and their expected split upon the introduction of an analyte can be used to increase sensitivity of biosensors based on WGM resonators.

**Acknowledgments**

This research was supported by ICore: the Israeli Center for Research Excellence "Circle of Light" grant no. 1902/12, and by the Israel Science Foundation grant no. 2013/15. Lev Deych and Vladimir Shuvayev acknowledge partial financial support of PSC-CUNY via grant #67388-00 45.